# Sub-Volt Silicon-Organic Electrooptic Modulator

Ran Ding[1,*], Tom Baehr-Jones[1,*], Woo-Joong Kim[2], Alexander Spott[1], Jean-Marc Fedeli[3], Su Huang[4], Jingdong Luo[4], Alex K.-Y. Jen[4], Larry Dalton[5], and Michael Hochberg[1]

[1]Department of Electrical Engineering, University of Washington, Campus Box 352500, Seattle, WA 98195, USA

[2]Department of Physics, Seattle University, 901 12th Ave., Seattle, WA, 98122, USA

[3]CEA LETI, Minatec 17 rue des Martyrs, 38054 Grenoble, France

[4]Department of Materials Science and Engineering, Campus Box 352120, University of Washington, Seattle, WA 98195, USA

[5]Department of Chemistry, University of Washington, 109 Bagley Hall Box 351700, Seattle WA 98195, USA

[*] Address correspondence to dingran@uw.edu (R. D.), or to baehrjt@washington.edu (T. B.-J.)

**Lowering the operating voltage of electrooptic modulators is desirable for a variety of applications, most notably in analog photonics[1,2] and digital data communications[3]. In particular for digital systems such as CPUs, it is desirable to develop modulators that are both temperature-insensitive and compatible with typically sub-2V CMOS electronics[4]; however, drive voltages in silicon-based MZIs currently exceed 6.5V[5]. Here we show an MZI modulator based on an electrooptic polymer-clad silicon slot waveguide, with a halfwave voltage of only 0.69V, and a bandwidth of 500 MHz. We also show that there are also paths to significantly improve both the bandwidth and drive voltage[6]. Our silicon-organic modulator has an intrinsic power consumption less than 0.66 pJ/bit, nearly an order of magnitude improvement over the previous lowest energy silicon MZI[7].**

The first silicon MZI modulator was based on carrier-depletion in a p-n structure[8], with bandwidth around 1 GHz and a modulation figure of merit of 8 V-cm. Improvements in design have led to silicon MZI modulators that exhibit modulation figures of merit of 4 V-cm at 30 GHz[9], and more recently 1.4 V-cm at 12 GHz[10]. But in the latter case, the authors identify a fundamental tradeoff between carrier concentration and modulator performance; that figure of merit can only be obtained with around 19 dB/cm of intrinsic absorption loss. It may never be possible to build a practical MZI modulator in silicon based on carrier depletion with a halfwave voltage of less than 1 Volt. A similar limitation has been encountered with Lithium Niobate based modulators; at speeds near 20 GHz, halfwave voltages are typically near 2.7 V or higher, while even at speeds as low as 1 GHz, halfwave voltages are still 1.2V or more[11,12]. Typical Lithium Niobate device lengths for these halfwave voltages are 5 cm or longer. With forward-biased diode based silicon MZIs[7], the halfwave voltage can be lowered to 1.8V, but the actual drive voltages for RF bitrates are around 7.6V[5]. Resonant enhancement[13], and electroabsorption modulators[14] can reduce drive voltages, but introduce other limitations, and would be unsuitable for use in analog links[15] or as phase modulators[1,2].

Slot waveguides were first proposed as a means to focus the propagating optical mode outside the silicon[16]. Later, it was demonstrated that by electrically contacting both arms of a slot waveguide, and coating the waveguide with an electrooptic polymer[17] cladding, a particularly responsive modulator could be created[18]; in this device, the performance of the organic active material was enhanced by the RF and optical mode confinement of the slot waveguide. Slot-waveguide polymer MZI modulators have since been demonstrated with halfwave voltages of 0.25V[19], though only at quasi-DC speeds. Another approach combines an electrooptic polymer-clad slot waveguide with a photonic crystal[20,21,22]. We have recently demonstrated that slot-waveguide polymer MZI modulators can work at RF speeds[23]. Here, we demonstrate, for the first time, a silicon-organic modulator with a low absolute drive voltage at RF speeds. A 9 mm MZI modulator results in a halfwave voltage of 0.69V with a bandwidth of 500 MHz; this corresponds to a modulator figure of merit of 0.61 V-cm. The modulator reported here is the first silicon photonic device to demonstrate an absolute operating voltage compatible with the low operating voltage of modern CPUs: The next-best result operates at 6.5V[5], almost an order-of-magnitude worse than this result.

Silicon waveguides were fabricated with two steps of self-aligned photolithography with a 193 nm stepper and dry etching. The initial substrate was SOI, with a 220 nm thick silicon layer on a 2000 nm oxide layer, on a silicon handle with resistivity of 10Ω-cm. The wafer was diced, and further fabrication and testing occurred on individual dies. A blanket implant was performed, with a target concentration of $3 \times 10^{17}$ cm$^{-3}$ phosphorous across the entire chip. A masked implant with final concentration target $10^{20}$ cm$^{-3}$ phosphorous was done for pad contacts, with contact photolithography, followed by metallization with a 10 μm clearance from the slot waveguide to the metal. The 10 μm clearance was used to ease alignment tolerances, and could be significantly decreased in future devices. The metal layer consisted of 2 μm of aluminium topped with 10 nm of gold. Final device layout, illustrated in Figure 1, consisted of a 9 mm MZI, formed with a strip-loaded slot waveguide and two 10 μm wide metal contacts, optically coupled via grating couplers[24]. The slot size was 200nm, and the two sides of the slot waveguide had widths 230 nm and height 220 nm, while the strip-loaded height was 68 nm. Figure 2 shows the geometry and mode pattern of the contacted strip-loaded slot waveguide.

A path length difference of 80 µm for the two MZI arms is used; this allowed the phase shift to be measured by device transmission, as shown in the methods, and enabled setting the MZI bias point by tuning the wavelength. The device dimensions are approximately 9 mm x 200 µm excluding contact pads, resulting in a total area of 1.8 mm$^2$. The device layout was driven by the need to manually probe between the waveguides, which requires large pads. Considering only the metal leads and silicon waveguide and excluding the probe pads, the total area is 0.8 mm$^2$.

The device was spin-coated (1500 rpm) with an electrooptic polymer that consisted of PMMA doped with 14 wt% of AJLZ53 chromophore[25] in chlorobenzene, leaving a film of around 2 µm. The device was then baked at 75 °C under vacuum. This electrooptic polymer typically exhibits a refractive index around 1.53, a relative dielectric constant of 3.2 to 3.5 at RF frequencies, and optical loss on the order of 1 dB/cm for wavelengths near 1550 nm. The device was poled in push-pull configuration, with the three pads biased at –20V, ground and 20V, with a poling temperature of around 110C.

Measurements of the device were initially taken with an Agilent 81980A tunable laser, and an Agilent 81636B fast power sensor. The insertion loss of the device was measured to be, at peak transmission, around 50 dB fiber to fiber. This includes approximately 9 dB per grating coupler and 2 dB from other fiber connections, leaving 30 dB for the device. Testing was typically performed with 13 dBm of laser power. Due to simulations we attribute approximately 8 dB to the two y-junctions and ridge to slot converters, leaving 22 dB of loss from the 9 mm of strip-loaded slot waveguide, and just over 1 cm of the connecting ridge waveguide. Much of this loss is is caused by imperfections in waveguide fabrication. Unimplanted strip-loaded slot waveguides have been built with 6.5 dB/cm of loss[26]. By using graded implants, we have shown that it will be possible to significantly decrease the losses of these modulators in the future[15].

To characterize modulator performance at low speeds, optical transmission spectra with several different DC bias voltages were taken, showing a fringe spacing of around 6.9 nm due to the arm length imbalance. Figure 3 shows the device transmission spectrum under two different bias voltages, as well as the phase shift deduced from the location of one of the peaks. The slight increase in insertion loss at shorter wavelengths is due to the bandwidth of the grating coupler. The DC device resistance was high, typically in excess of 100 kΩ. A halfwave voltage of 0.69±0.07V was derived as the average of the two measurement sets. When combined with a modulation figure of merit for the waveguide of 0.21 μm$^{-1}$, this suggests an $r_{33}$ value for the poled polymer of around 54 pm/V[27]. This is slightly lower than the peak performance for the material of 65 pm/V, which we confirmed from a measurement of a single layer film on ITO.

To characterize the device at RF speeds, the output from the device was connected to a New Focus 1647 avalanche photodetector with 1 GHz bandwidth, and an S-parameter was taken with an Agilent E8361C network analyzer. The wavelength was chosen to bias the device at the 3dB point. Figure 4 shows the predicted S21 value for the DC halfwave voltage, along with the measured S21 value. The modulation bandwidth, typically defined as the 6 dB point for an RF s-parameter[8,23], is around 500 MHz. The s-parameter was measured with -10 dBm RF power, to ensure that the modulator remained in the linear regime.

The bandwidth limitation seen is almost certainly not due to the electrooptic polymer; similar polymers have shown bandwidths of 165 GHz[28]. Nor is it likely that carrier transit times are the source of the limitation; the silicon acts only as a transparent conductor. The likely explanation is parasitic capacitance and resistance. The equivalent circuit for both arms of the device is shown in Figure 4. The resistance due to the strip-loaded arms, $R_{sl}$, should be around 8Ω. From a finite element simulation, $C_{sl}$ should be approximately 1.8 pF. $C_s$ should be 3.1 pF, while $R_s$ can be calculated as 6Ω. Using these values, a 6dB rolloff in S21 should be at 780 MHz. But the S11 would only be -0.8 dB in this case. Most likely $R_{sl}$ is larger than anticipated; at 75Ω, a 6dB rolloff at around 500 MHz is predicted with an S11 value of -3 dB, in agreement with the experimental data. Etch-induced damage on the relatively thin strip-loaded region may explain the higher resistance[29]. Device speed could be improved by decreasing metal clearance, thus lowering $R_{sl}$, or through a higher implant dose. High resistivity substrate wafers, which are available commercially from SOITEC, could also decrease $R_s$.

The intrinsic device power consumption will be driven by the energy required to charge the capacitors in the device equivalent circuit, assuming that the parasitic resistances can be lowered. Conservatively attributing all of the S21 rolloff to capacitive loading, total device capacitance is 11 pF, significantly higher than the previously calculated value. In reality, some of the rolloff is due to resistive loading as well, as exhibited in the S11 parameter. With a value of 11 pF, and assuming that in a random NRZ-encoded bitstream the device must be charged only every fourth bit on average[13], the intrinsic energy consumption will be at most 0.66 pJ/bit. This compares favorably to the best results in silicon, thus far, which are 5pJ/bit at 10 GBit[7]. Similarly, we calculate that the best result for a lithium niobate modulator at 1 GBit/s would be 7.2 pJ/bit[Error! Bookmark not defined.], but this would likely require a device length in excess of 5 cm. We note that the 500 MHz bandwidth shown is nearly sufficient for 1 GBit operation.

To conclude, we have demonstrated a MZI modulator with a sub-1V halfwave voltage at RF bandwidth. This is to our knowledge the lowest halfwave voltage shown for any silicon-based MZI, and appears to be the lowest halfwave voltage shown in any material system for an RF electrooptic MZI modulator. Moreover, there is a clear path towards improved performance, from both better RF design[15] and more active polymers; in particular, polymers with $r_{33}$ values of 300 pm/V have been demonstrated[30], which, if used in our device, could result in a halfwave voltage of 120 mV. Narrower slots would also improve performance[6]. We believe that slot-waveguide polymer modulators will play an important role in future silicon photonic systems.

**Methods**

Here we present a derivation of the methodology used to measure the halfwave voltage with device transmission. To a constant factor, the transmission through an unbalanced MZI can be expressed as a function of wavelength as

$$\frac{1}{2}\left(1+\cos\left(\frac{2\pi}{\lambda}n(\lambda)\Delta L \pm \frac{\pi V}{V_\pi}\right)\right) \qquad (1)$$

where $\lambda$ is the wavelength, $n(\lambda)$ is the effective index, which is generally a function of wavelength, $\Delta L$ is the length difference between the arms, $V$ is the bias voltage, and finally $V\pi$ is the halfwave voltage. The sign in front of the bias term will be determined by the polarity of the device poling, in combination with the orientation of the $\chi^2$ moment with respect to the poling field. One can expand the argument to the cosine in (1) to first order in $\lambda$ as

$$\frac{2\pi\Delta L n(\lambda_0)}{\lambda_0} - \frac{2\pi(\lambda-\lambda_0)}{\lambda_0}n_g\frac{\Delta L}{\lambda_0} \pm \frac{\pi V}{V_\pi} \qquad (2)$$

where $\lambda_0$ for convenience is chosen to be a wavelength where for no bias, the device has a minimum in transmission. This amounts to neglecting, among other things, group velocity dispersion. The minimum in the transmission spectra will be found where (2) is equal to $\pi$ radians. It then is immediately clear that the shift in the location of a transmission minimum will have a linear relationship to both the bias voltage and the induced phase shift. The phase shift can be expressed most conveniently in terms of the fringe to fringe spacing $\Delta_\lambda$.

$$\pm\frac{\pi V}{V_\pi} = \Delta\phi = \frac{2\pi(\lambda - \lambda_0)}{\lambda_0} n_g \frac{\Delta L}{\lambda_0} = \frac{2\pi}{\Delta_\lambda}(\lambda - \lambda_0) \qquad (3)$$

To determine the location of a minimum in the transmission spectra, we selected the very lowest power level in a given region, and then averaged all points within 3 dB of this level. The variance of this distribution was then used as an uncertainty in the measurement.

**Figures**

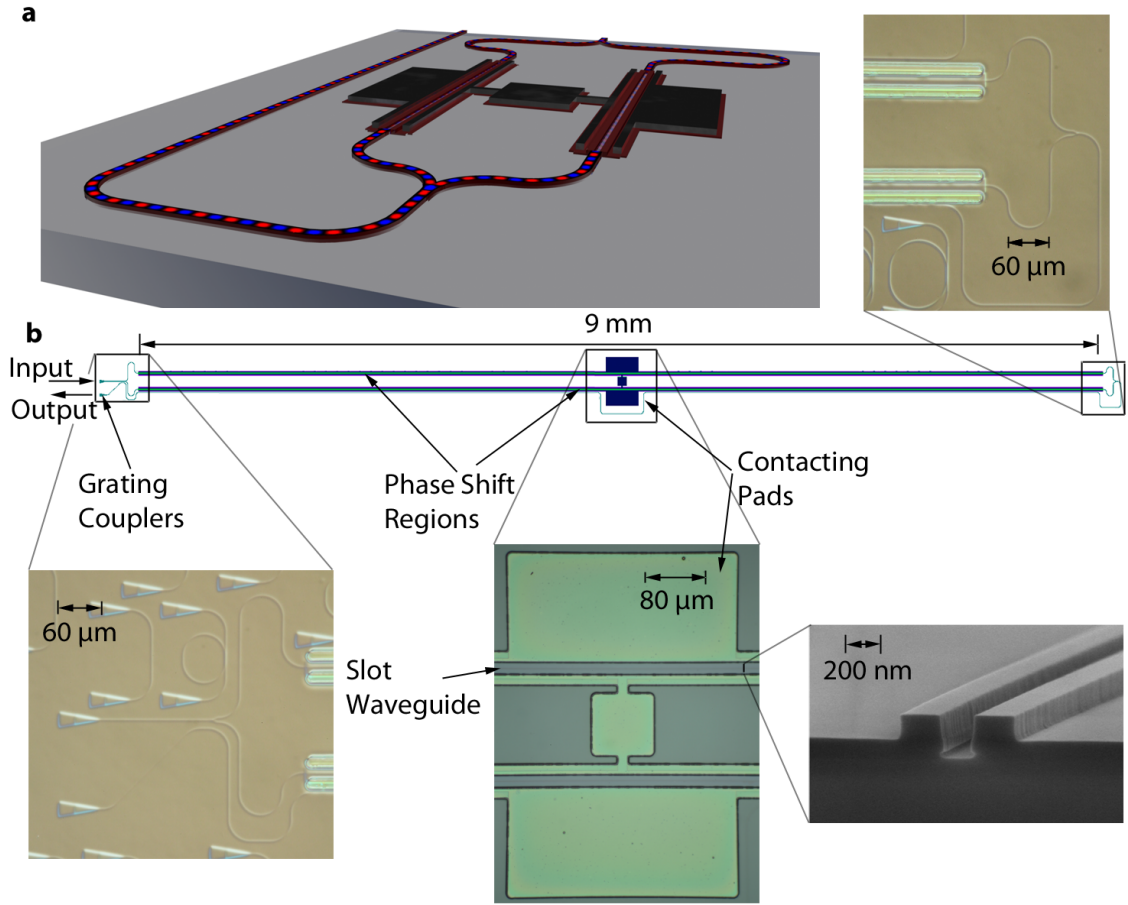

**Figure 1 Device Layout. a,** A rendering of a portion of the MZI is shown, illustrating the optical path. **b,** The layout of the device is shown. Note that after traveling through the MZI, the optical signal returns back to the side of the MZI in a ridge waveguide. Several optical micrographs and an SEM micrograph of key features are also displayed. It should be noted that in two of the optical micrographs, portions of adjacent, unrelated devices are shown.

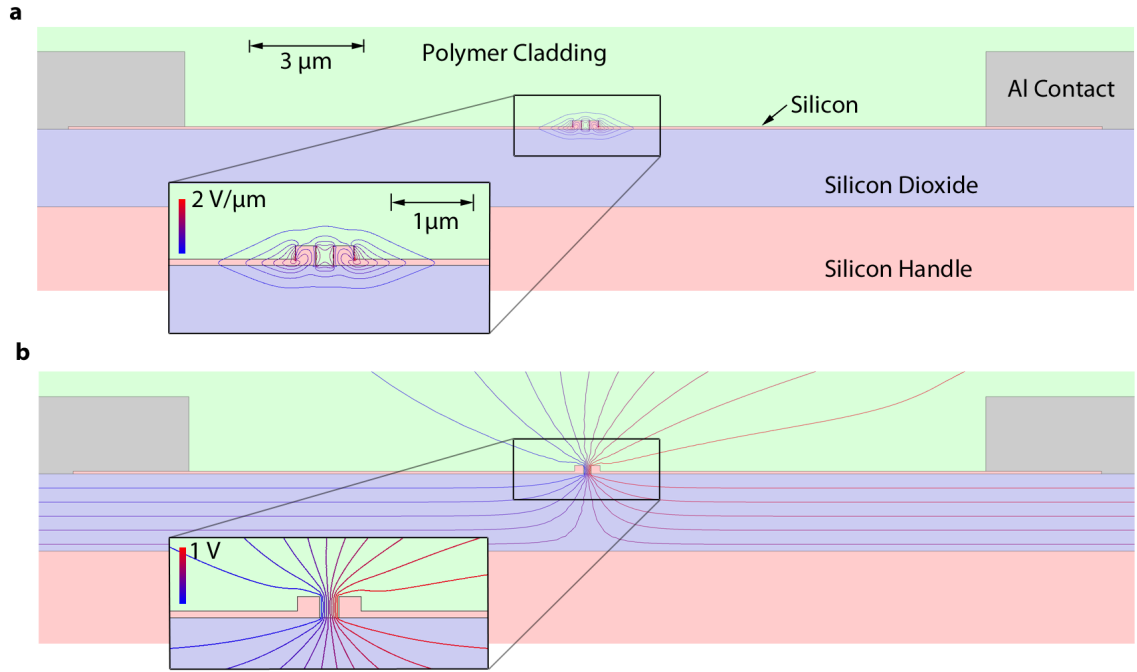

**Figure 2 Device Layout and Modal Structure. a**, The contacted strip-loaded slot structure is shown, as well as the optical mode. A contour plot of $|E_x|$ is shown, normalized for 1 mW of propagating power. **b**, The RF mode is shown, with a contour plot of the voltage that would be seen at high frequencies, but below the bandwidth limit.

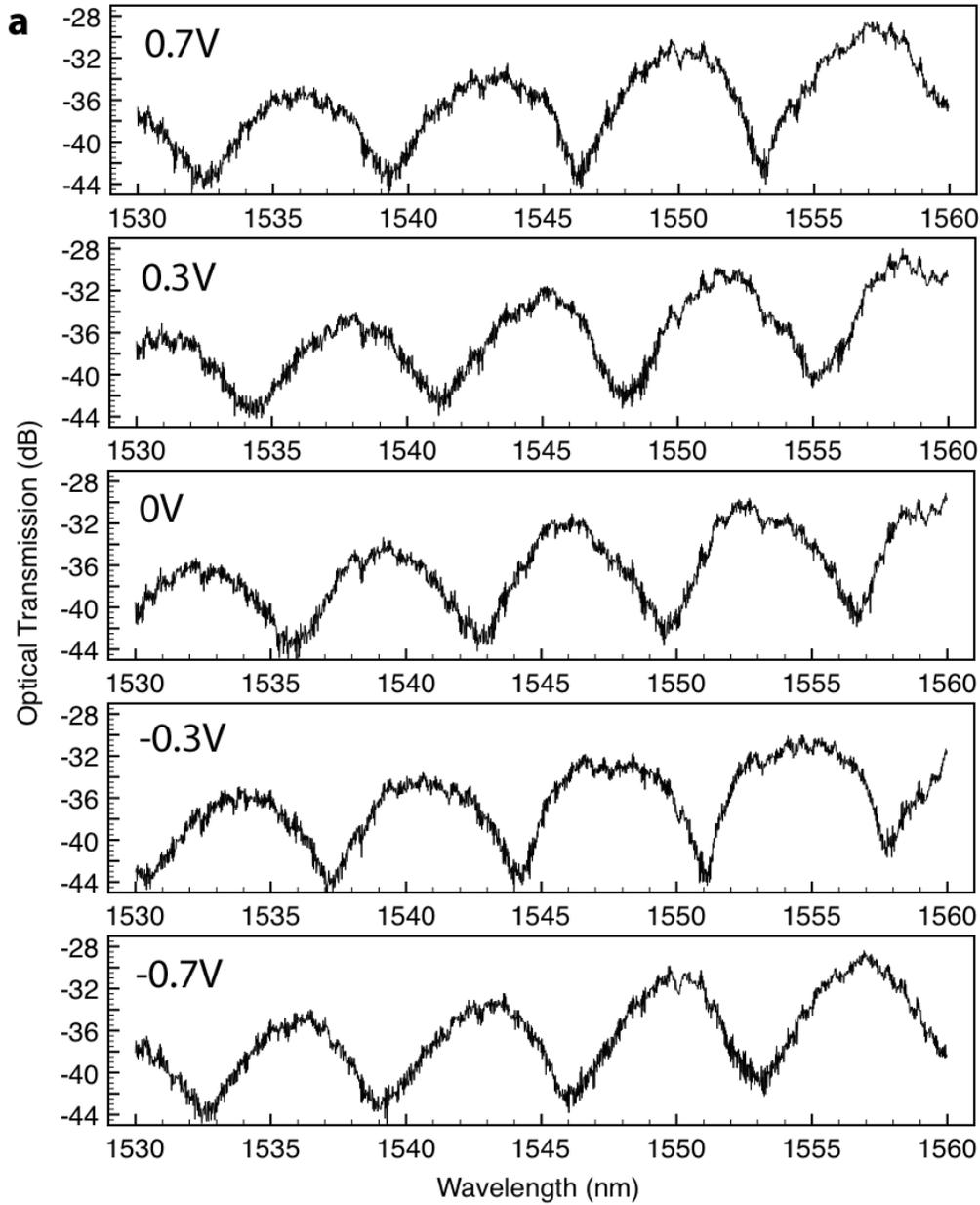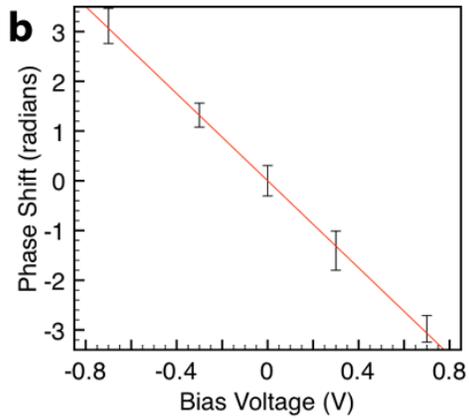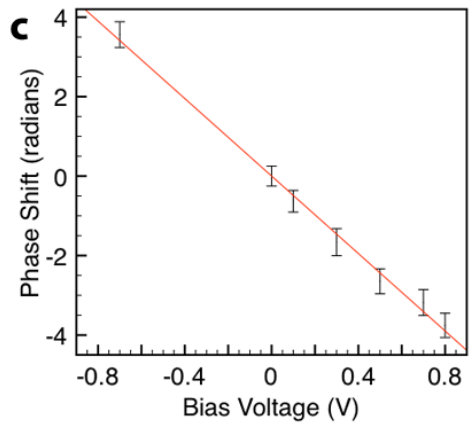

**Figure 3 DC device performance. a,** Device transmission excluding off-chip coupling losses as a function of wavelength is shown for several bias voltages. **b, c,** A plot of phase shift in radians as a function of bias voltage for two different measurement sets, determined by the migration of the minimum near 1550 nm as a function of bias voltage. The slopes indicate halfwave voltages of 0.72±0.06V and 0.65±0.03V, respectively. The device transmission plots correspond to the data points of the first measurement set.

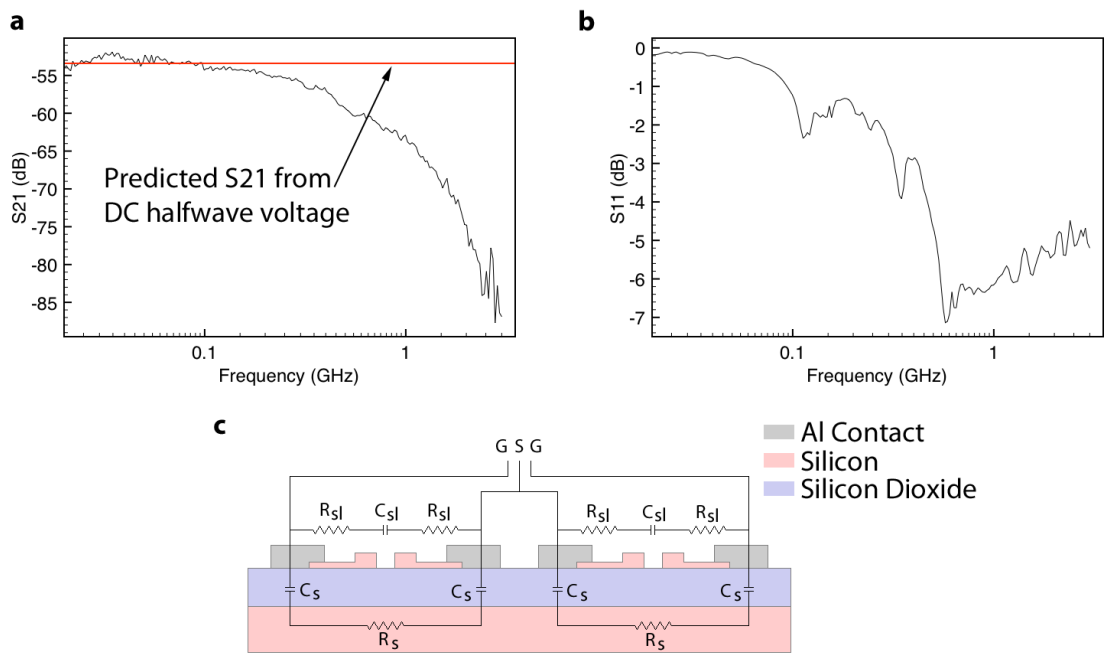

**Figure 4 RF performance and source of bandwidth limit. a,** The measured S21 parameter is shown as a function of frequency, along with the predicted value from the DC results based on the known losses of the system and photodetector conversion gain.

**b**, The S11 parameter is shown, with a 3 dB rolloff at around 500 MHz. **c,** A rendering of the equivalent circuit of the device is shown.

[1] Chou, H.-F. *et al.* Highly Linear Coherent Receiver With Feedback. *IEEE Photonics Technology Letters* **19**, 940-942 (2007).
[2] Macario, J. *et al.* Development of Electro-Optic Phase Modulator for 94 GHz Imaging System. *Journal of Lightwave Technology* **27**, 5698-5703 (2009).
[3] Lipson, M. Silicon photonics: the optical spice rack. *Electronics Letters* **45**, 8-10 (2009).
[4] Jaeger, R. C. & Blalock, T. N. *Microelectronic Circuit Design* (McGraw Hill, 2008).
[5] Reed, G. T., Mashanovich, G., Gardes, F. Y. & Thomson, D. J. Silicon optical modulators. *Nature Photonics* **4**, 518-526 (2010).
[6] Hochberg, M. *et al.* Towards a millivolt optical modulator with nano-slot waveguides. *Optics Express* **15**, 8401-8410 (2007).
[7] Green, W. M. J., Rooks, M. J., Sekaric, L., & Vlasov, Y. A. Ultra-compact, low RF power, 10 Gb/s silicon Mach-Zehnder modulator. *Optics Express* **15**, 17106-17113 (2007).
[8] Liu, A. *et al.* A high-speed silicon optical modulator based on a metal–oxide–semiconductor capacitor. *Nature* **427**, 615-618 (2004).
[9] Liao, L. *et al.* 40 Gbit/s silicon optical modulator for highspeed applications. *Electronics Letters* **43**, 1196-1197 (2007).
[10] Feng, N.-N. *et al.* High speed carrier-depletion modulators with 1.4V-cm V$\pi$L integrated on 0.25μm silicon-on-insulator waveguides. *Optics Express* **18**, 7994-7999 (2010).
[11] PhotonicSystems, part number PSI-3600-MOD-D1.
[12] Aoki, K. *et al.* Low Half-Wave Voltage X-Cut Thin LiNbO$_3$ Sheet Optical Phase Modulator With Asymmetric Coplanar Waveguide Electrode. *IEEE Photonics Technology Letters* **20**, 1811-1813 (2008).
[13] Dong, P. *et al.* Low V$_{pp}$, ultralow-energy, compact, high-speed silicon electro-optic modulator. *Optics Express* **17**, 22484-22490 (2009).
[14] Liu, J. *et al.* Waveguide-integrated, ultralow-energy GeSi electro-absorption modulators. *Nature Photonics* **2**, 433-437 (2008).
[15] Witzens, J., Baehr-Jones, T. & Hochberg, M. Design of transmission line driven slot waveguide Mach-Zehnder interferometers and application to analog optical links. *Optics Express* **18**, 16902-16928 (2010).
[16] Almeida, V. R., Xu, Q. F., Barrios, C. A. & Lipson, M. Guiding and confining light in void nanostructure. *Optics Letters* **29**, 1209-1211 (2004).
[17] Enami, Y. *et al.* Hybrid cross-linkable polymer/sol-gel waveguide modulators with 0.65 V half wave voltage at 1550 nm. *Applied Physics Letters* **91**, 093505 (2007).
[18] Baehr-Jones, T. *et al.* Optical modulation and detection in slotted silicon waveguides. *Optics Express* **13**, 5216-5226 (2005).

[19] Baehr-Jones, T. *et al.* Nonlinear polymer-clad silicon slot waveguide modulator with a half wave voltage of 0.25 V. *Applied Physics Letters* **92**, 163303 (2008).
[20] Loncar, M. *et al.* Waveguiding in planar photonic crystals. *Applied Physics Letters* **77**, 1937-1939 (2000).
[21] Leuthold, J. *et al.* Silicon Organic Hybrid Technology – A Platform for Practical Nonlinear Optics. *Proceedings of the IEEE* **97**, 1304-1316 (2009).
[22] Wülbern, J. H. *et al.* Electro-optic modulation in slotted resonant photonic crystal heterostructures. *Applied Physics Letters* **94**, 241107 (2009).
[23] Ding, R. *et al.* Demonstration of a low $V_\pi L$ modulator with GHz bandwidth based on electro-optic polymer-clad silicon slot waveguides. *Optics Express* **18**, 15618-15623 (2010).
[24] Roelkens, G., Van Thourhout, D., & Baets, R. High efficiency Silicon-on-Insulator grating coupler based on a poly-Silicon overlay. *Optics Express* **14**, 11622-11630 (2006).
[25] Greenlee, C. *et al.* Mach–Zehnder interferometry method for decoupling electro-optic and piezoelectric effects in poled polymer films. *Applied Physics Letters* **97**, 041109 (2010).
[26] Ding. R, *et al.* Low-loss Strip-Loaded Slot Waveguides in Silicon-on-Insulator. *Optics Express*, under consideration.
[27] Hochberg, M. *et al.* Towards a millivolt optical modulator with nano-slot waveguides. *Optics Express* **15**, 8401-8410 (2007).
[28] Bortnik, B. *et al.* Electrooptic Polymer Ring Resonator Modulation up to 165 GHz. IEEE Journal of Selected Topics in Quantum Electronics **13**, 104-110 (2007).
[29] Streetman, B. G. *Solid State Electronics Devices* (Prentice Hall, 1995).
[30] Huang, S. *et al.* Highly efficient electro-optic polymers through improved poling using a thin TiO2-modified transparent electrode. *Applied Physics Letters* **96**, 243311 (2010).